%% file: main.tex
\documentclass[aps, prb, superscriptaddress, twocolumn, reprint]{revtex4-2}
\usepackage{lineno}  
\usepackage[caption=false]{subfig}
\usepackage{graphicx}
\usepackage{multirow}
\usepackage{array}

\usepackage[page]{appendix}

\usepackage{hyperref}

\usepackage{amsmath}

\usepackage{xcolor}

\usepackage{tikz}
\usetikzlibrary{shapes.geometric, arrows}

\usepackage{varwidth}

\usepackage[capitalize]{cleveref}
\crefname{section}{Sec.}{Secs.}
\Crefname{section}{Section}{Sections}

\usepackage{mathrsfs}
\usepackage{dsfont}

\usepackage{transparent}

\newif\ifnightmode
\newif\ifincludeinternal

\input{common_commands}

\ifnightmode
    \pagecolor[rgb]{0.1,0.1,0.1}
    \color[rgb]{0.77,0.77,0.77}
    \newcommand{\transpcmd}{\transparent{0.4}}
\else
    \newcommand{\transpcmd}{}
\fi

\begin{document}
\title{Predicting ionic conductivity in solids from the machine-learned potential energy landscape}

\newcommand{\aflNUSIFIM}{%
    National University of Singapore, Institute for Functional Intelligent Materials,
    NUS S9 Building, 4 Science Drive 2, 117544 Singapore
}
\newcommand{\aflNUSYALE}{%
    Yale-NUS College,
    16 College Avenue West, 138527 Singapore
}
\newcommand{\aflNUSCATDM}{%
    National University of Singapore, Centre for Advanced 2D Materials,
    6 Science Drive 2, 117546 Singapore
}
\newcommand{\aflNUSMSE}{%
    National University of Singapore, Department of Materials Science Engineering,
    9 Engineering Drive 1, 117575 Singapore
}
\newcommand{\aflHSE}{%
    HSE University, Faculty of Computer Science,
    Pokrovsky Boulevard 11, 109028 Moscow, Russian Federation
}
\newcommand{\aflCONSTRUCTOR}{%
    Constructor University,
    Bremen, Campus Ring 1, 28759, Germany
}
\newcommand{\aflNUSPHY}{%
    National University of Singapore, Department of Physics,
    2 Science Drive 3, 117551 Singapore
}
\newcommand{\aflCONSTRUCTORKL}{%
    Constructor Knowledge Labs,
    Bremen, Campus Ring 1, 28759, Germany
}

\author{Artem Maevskiy}\email[Contact author: ]{maevskiy@nus.edu.sg}\affiliation{\aflNUSIFIM}
\author{Alexandra Carvalho}\affiliation{\aflNUSIFIM}\affiliation{\aflNUSCATDM}
\author{Emil Sataev}\affiliation{\aflHSE}
\author{Volha Turchyna}\affiliation{\aflHSE}
\author{Keian Noori}\affiliation{\aflNUSIFIM}\affiliation{\aflNUSCATDM}
\author{Aleksandr Rodin}\affiliation{\aflNUSYALE}\affiliation{\aflNUSMSE}
\author{A.\,H.\,Castro Neto}\affiliation{\aflNUSIFIM}\affiliation{\aflNUSCATDM}\affiliation{\aflNUSMSE}\affiliation{\aflNUSPHY}
\author{Andrey Ustyuzhanin}\affiliation{\aflCONSTRUCTORKL}\affiliation{\aflCONSTRUCTOR}\affiliation{\aflNUSIFIM}


\begin{abstract}
    Discovering new superionic materials is essential for advancing solid-state
    batteries, which offer improved energy density and safety compared to the
    traditional lithium-ion batteries with liquid electrolytes. Conventional computational
    methods for identifying such materials are resource-intensive and not
    easily scalable.
    Recently, universal interatomic potential models have been developed
    using equivariant graph neural networks. These models are trained on extensive datasets of
    first-principles force and energy calculations. One can achieve significant computational advantages
    by leveraging them as the foundation for
    traditional methods of assessing the ionic conductivity, such as molecular dynamics or
    nudged elastic band techniques. However, the
    generalization error from model inference on diverse atomic structures arising in
    such calculations can compromise the reliability of the results. In this work, we
    propose an approach for the quick and reliable screening of ionic conductors through
    the analysis of a universal interatomic potential. Our method incorporates
    a set of heuristic structure descriptors that effectively employ the rich knowledge of the underlying
    model while requiring minimal generalization capabilities. Using our descriptors,
    we rank lithium-containing materials in the Materials Project database according to their
    expected ionic conductivity. Eight out of the ten highest-ranked materials are confirmed
    to be superionic at room temperature in first-principles calculations. Notably, our method achieves a speed-up factor
    of approximately 50 compared to molecular dynamics driven by a machine-learning potential, and is
    at least 3,000 times faster compared to first-principles molecular dynamics.
\end{abstract}

\maketitle

\ifnightmode\color[rgb]{0.77,0.77,0.77}\fi
\section{Introduction}

Solid-state batteries (SSBs) are a promising alternative to traditional lithium-ion batteries
due to their higher energy density and enhanced safety features. With the elimination of liquid
electrolytes, SSBs reduce the risk of leakage and combustion, thus mitigating one of the major
safety concerns with conventional batteries~\cite{kim2015review,li2021advance,janek2023challenges}.
Coupled with the potential for increased energy storage~\cite{dudney2015handbook},
this safety enhancement makes SSBs a crucial technology for advancing electric vehicles and portable
electronics, where the demand for longer battery life and robust performance is ever-increasing.

Despite these advantages, SSBs development faces a number of challenges that prevents their widespread
adoption~\cite{Zhao2020}. One major limitation is the relatively low
ionic conductivity of solid electrolytes compared
to their liquid counterparts, which can diminish the overall performance of the battery. Additionally,
interface stability between the solid electrolyte and electrodes remains a critical
issue~\cite{chen2021interface,ohta2006enhancement}. The formation
of interfacial resistance and degradation over time can significantly impact the lifespan and efficiency
of a battery~\cite{chen2021interface}. Therefore, there is a pressing need for
the development of new solid-state electrolytes (SSEs)
with improved properties to overcome these barriers.

To address these challenges, computational prediction methods have been employed to identify and
optimize new SSE materials~\cite{PhysRevB.105.224310, Carvalho2022, C9EE02457C, muy2019high, wang2015design}.
However, these methods are often computationally expensive and time-consuming.
Machine learning (ML) presents a viable solution to mitigate these limitations~\cite{schutt2017schnet,
xie2018crystal, chen2019graph, chen2022universal, Deng2023, Batatia2022Design, Batatia2022mace, Dunn2020,
batatia2023foundation, Merchant2023, zeni2024mattergen, liu2021recent, hu2022smart}. By leveraging large
datasets such as the Materials Project database~\cite{jain2013commentary} and advanced algorithms,
ML has the potential to speed up the discovery of new materials with
high ionic conductivity and therefore accelerate the research and development of high-performance
SSBs~\cite{laskowski2023identification, sendek2020quantifying}.

Different ML algorithms that have been employed to assist the discovery of superionic
materials include classical supervised ML methods like logistic
regression~\cite{sendek2017holistic,sendek2018machine},
gradient boosting~\cite{choi2021searching} and support vector
machines~\cite{fujimura2013accelerated, cubuk2019screening}, classical unsupervised
methods~\cite{zhang2019unsupervised, laskowski2023identification}, as well as deep learning
approaches based on graph neural networks (GNN)~\cite{ahmad2018machine, Wang2023EndToEnd,
guo2024machine} and transformers~\cite{Hargreaves2023}.
The main difficulty in implementing the ML models for identifying SSEs is in general the limited
availability of high-quality data on ionic mobility in solids. This motivates the use of
knowledge transfer
methods~\cite{cubuk2019screening}, or trading off for larger datasets~\cite{Wang2023EndToEnd} obtained
using approximate methods~\cite{NISHITANI2018111}.

The scarcity of available data drives the search for informative material descriptors that
correlate well with ionic conductivity.
This search is typically focused on descriptors that
characterize compositions, geometric configurations, and electrochemical and electronic properties of
materials~\cite{Jang2022Na, laskowski2023identification, he2020high}, often relying on implementations
from the \texttt{matminer} library~\cite{WARD201860}, which are
adapted from scientific publications. In this work, we aim to expand this focus by exploring the
features of the interatomic potential (IAP).

While IAP can be evaluated with expensive first-principles density functional theory (DFT) calculations,
ML models allow for bypassing this step. Since there is notably more high-quality data available
for predicting interatomic forces and energies, the task of predicting IAP is far less constrained
than identifying SSEs directly. In fact, various ML methods have been
proposed in the past decades to effectively bypass DFT in IAP calculations~\cite{behler2007generalized,
bartok2010gaussian, thompson2015spectral}.

A notable improvement in the performance and reliability of these methods has been observed after the
adoption of graph neural networks (GNNs)~\cite{schutt2017schnet, xie2018crystal, chen2019graph}.
These models inherently impose permutation invariance by operating
on graph structures, where the order of nodes does not
affect the computation of graph-level representations. Additionally, invariance under 3D translations
and rotations is achieved through the use of equivariant message passing and invariant feature
aggregation methods.
Although most of the initial models utilize 2-body messages, the introduction of
many-body messages and updates allowed models like CHGNet~\cite{Deng2023},
M3GNet~\cite{chen2022universal}, MACE~\cite{Batatia2022Design,Batatia2022mace,batatia2023foundation}
and the NequIP-based~\cite{NequIP} SevenNet~\cite{sevennet}
achieve significant quality improvement, as demonstrated on the Matbench Discovery
public benchmark~\cite{riebesell2023matbench}. These models may also be called
universal potentials, as they cover the majority of the elements in the
periodic table that are highly relevant to applications.

In the context of ionic transport studies, ML-IAPs can serve as substitutes
for DFT in MD~\cite{guo2024machine} and
nudged elastic band (NEB) calculations \cite{jonsson1998nudged}.
However, a notable concern with
ML-IAPs is the generalization error, which refers to the error in the model's predictions
when applied to new data, unseen at training time. Naturally, one should expect this error to increase
as MD and NEB simulations drive the studied systems further from the domain of the training set,
leading to a decrease in the model's predictive accuracy and reliability.
A typical way of addressing
this issue is through active learning~\cite{Zeng_JChemPhys_2023_v159_p054801, Qi2024}.
This involves controlling the prediction error, running
DFT on configurations where the error is excessively large and then retraining the model
on these configurations. While this approach effectively reduces the generalization error,
it sacrifices some of the computational efficiency gained by switching from DFT to ML-IAPs.

In this work, we take a different approach by utilizing a universal potential model to
design a set of dedicated descriptors that predict materials with high lithium mobility while
minimizing generalization error.
We leverage an ML-IAP trained on vast amounts of data to analyze the potential landscape in a
controlled environment for the labeled structures present in high-quality computational and
experimental Li conductivity datasets.
From this analysis, we propose a set of heuristics that can be calculated on structure configurations
that are similar to the
examples used for training the potential. This ensures minimal generalization requirements for the
model and therefore more accurate predictions of the interatomic potential in the diverse material
environments.
Finally, we use these descriptors to predict SSEs within the Materials Project
database and validate our predictions with {\it ab initio} calculations.

The studies performed in this work are done
using the M3GNet model, which was trained by the authors of the original work~\cite{chen2022universal}
on the entire Materials Project dataset~\cite{jain2013commentary}. It should be noted, however, that
our approach is not specific to a particular model and could be used with any IAP.

This document is structured as follows:
\cref{l_met} describes the methodology used throughout this work, outlining the key ideas of the proposed
approach in \cref{l_met_pesff,l_met_val}, along with details of the {\it ab initio} calculations presented
in \cref{l_met_aimd}. In \cref{l_res}, we report and discuss our findings and provide an outlook. Finally,
we conclude in \cref{l_con}.

\section{Methodology}\label{l_met}

\begin{figure*}[htp]
    \centering
    {\transpcmd\includegraphics[width=\linewidth]{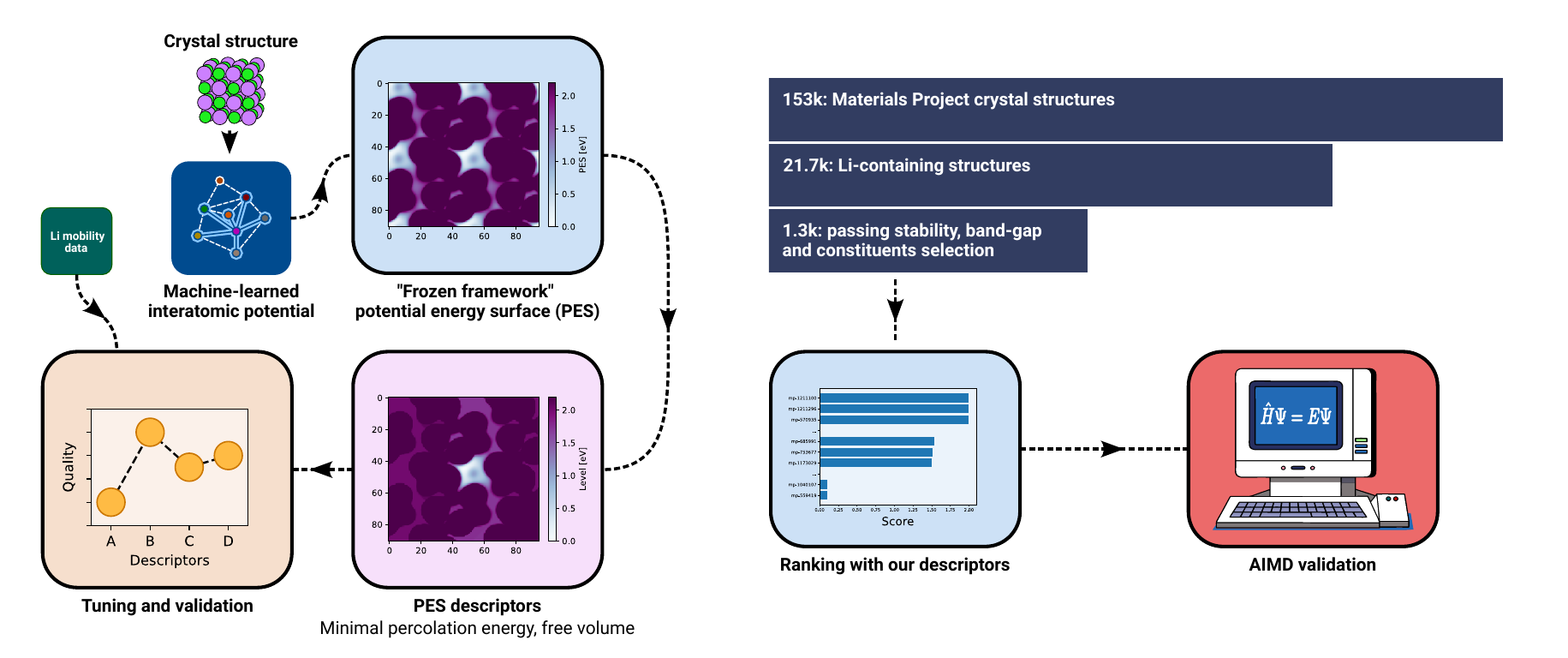}}
    \caption{\emph{Flowchart of the logical steps performed in our study}
    }
    \label{fig:flowchart}
\end{figure*}

The overall computational pipeline performed in this work is shown in \cref{fig:flowchart}.
We start from a machine-learned interatomic potential,
which is used to analyze the potential
energy surface (PES) for a given material. Then, a set of scalar heuristics is derived from the obtained
PES. These descriptors are based on our intuition of which PES properties should correlate well with ionic
mobility, and are inspired by the frozen host lattice idea behind the pinball model~\cite{PhysRevMaterials.2.065405}
and are similar to the pathway volume assessed with the bond-valence technique in~\cite{PhysRevLett.84.4144}.
We further validate our choice and evaluate the quality of the derived heuristics with
labeled data, both simulated and experimental. Descriptors that demonstrate robustness and highest
predictive power are chosen to rank the structures from the Materials Project
dataset~\cite{jain2013commentary}. Finally, we perform computationally demanding {\it ab initio}
molecular dynamics (AIMD) simulation to validate the candidates that are most promising according
to our method. We describe our pipeline in greater detail throughout
\cref{l_met_pesff,l_met_val,l_met_aimd}.

\subsection{Potential energy surface in the frozen framework approximation}\label{l_met_pesff}

In this work, we leverage the assumption that the generalization error of the ML-IAP model
is reduced when the system under study differs from the closest training object by only
the coordinates of a single atom, compared to a system where all the
atomic positions differ from those in the training set.
Therefore, each structure is considered as having a single mobile ion per unit
cell and a frozen framework formed by the remaining atoms. A potential
energy surface (PES) scan is performed by placing
the mobile ion at a set of locations
on a regular 3d grid and evaluating the potential energy for the resulting
structure. The grid spans over the entire
unit cell with axes parallel to the lattice vectors.
For each lattice vector $\vec{l}_i$, the grid step along that direction is
$\vec{g}_i = \vec{l}_i / N_i$, where $N_i$ is the smallest integer satisfying
$|\vec{l}_i| / N_i \leq 0.25$\,\AA.
Locations that are closer than $R_{\text{min}} = 1.2$\,\AA\ to any of
the other atoms are excluded from this process to avoid ML-IAP evaluations
on unphysical configurations.

Once the PES values on the grid are obtained, we repeat them along each
axis to create a $2\times2\times2$ supercell. We position the origin of
the grid so that the ion's lowest energy location is at the center of the
supercell, with its 26 replicas located at the outer faces. In other words,
the resulting PES is a 3-dimensional array with values corresponding to
the mobile ion locations $\vec{r}=\sum_i (o_i + c_i)\cdot\vec{g}_i$, with
$i\in\{1,2,3\}$, $o_i$ being the grid coordinates for the minimal PES
value and $c_i\in\{-N_i,\ldots,N_i\}$.
Then, a breadth-first search is performed for
a minimum-energy-barrier path from the center to any of the faces of the
supercell. \emph{Minimal percolation energy} (MPE) for this ion is reported as
the energy barrier value for the discovered path.

Due to the frozen framework approximation, MPE values are expected to be much larger
than the actual activation energies. In reality, ionic hopping in superionic conductors occurs
in a timescale of ns~\cite{morimoto2019microscopic, van2023re}, which allows for
relaxation of the remaining ions and reduces the activation energies. However, our
expectation is that MPE reflects the strength of the interactions that are overcome
during ionic hopping, and thus correlates with the ionic mobility.

The breadth-first search of the optimal path requires calculating a quantity,
which we denote
as the \emph{level map} (LM), given by:
\begin{linenomath*}\begin{equation}
\text{LM}(\vec{r}) \equiv \underset{C\in \mathscr{C}(\vec{r}, \vec{r}_0)}{\min}
\left[\underset{{\vec{r}\,}'\in C}{\max}E({\vec{r}\,}')\right],
\end{equation}\end{linenomath*}
where $\vec{r}_0$ is the initial location of the mobile atom at the relaxed state,
$\mathscr{C}(\vec{r}, \vec{r}_0)$ is the set of all paths connecting
$\vec{r}$ and $\vec{r}_0$, and $E({\vec{r}\,}')$ is PES value at ${\vec{r}\,}'$.
The meaning of LM is the minimal amount of energy an ion needs to be able to get to
$\vec{r}$ from its original site, within the frozen framework approximation.
In our calculation, as explained above, the $\min$ and $\max$ operations are
performed on a discrete grid. Using the level map concept, MPE can be defined
as:
\begin{linenomath*}\begin{equation}
\text{MPE} \equiv \underset{\vec{r}\in \mathscr{F}}{\min}\left[\text{LM}(\vec{r})\right],
\end{equation}\end{linenomath*}
where $\mathscr{F}$ denotes the set of points located at the faces of the supercell,
where the replicas of the original site
are located.

Finally, we introduce another family of
structure characteristics that we call
\emph{free volumes} (FV) by calculating the average volume of PES/LM below a certain
threshold $t$:
\begin{linenomath*}\begin{equation}
\text{FV}_{X}(t) = \frac{1}{V}
\int\limits_{\text{supercell}}\mathds{1}\left[X\vphantom{^0_0}(\vec{r}) < t\right]d\vec{r},
\end{equation}\end{linenomath*}
where $V$ denotes the full volume of the supercell, $\mathds{1}[\cdot]$ is the indicator
function and $X(\vec{r})$ can be either $E(\vec{r})$ or $\text{LM}(\vec{r})$,
in which case the corresponding
FV is called \emph{disconnected} or \emph{connected}, respectively. This naming choice reflects
that, by the definition of LM, any set of points satisfying $\text{LM}(\vec{r}) < T$ is a
connected set, while this may not necessarily hold for $E(\vec{r})$.

Since a unit cell may contain more than one potentially mobile atom, aggregation is performed
over atoms of the mobile species. For the case of MPE, the aggregation operation is just the
minimum value of individual MPEs over the mobile atoms. For FVs,
the aggregation is done by superimposing
the indicator function outputs for different mobile atoms and combining them with logical OR
operation prior to volume calculation.

\begin{figure}[htp]
    \subfloat[\label{fig:vesta:mp-1211296}Li(BH)$_6$]{%
        \centering
        {\transpcmd\includegraphics[width=0.48\linewidth,trim={11.5cm 5.6cm 14.5cm 3.6cm},clip]{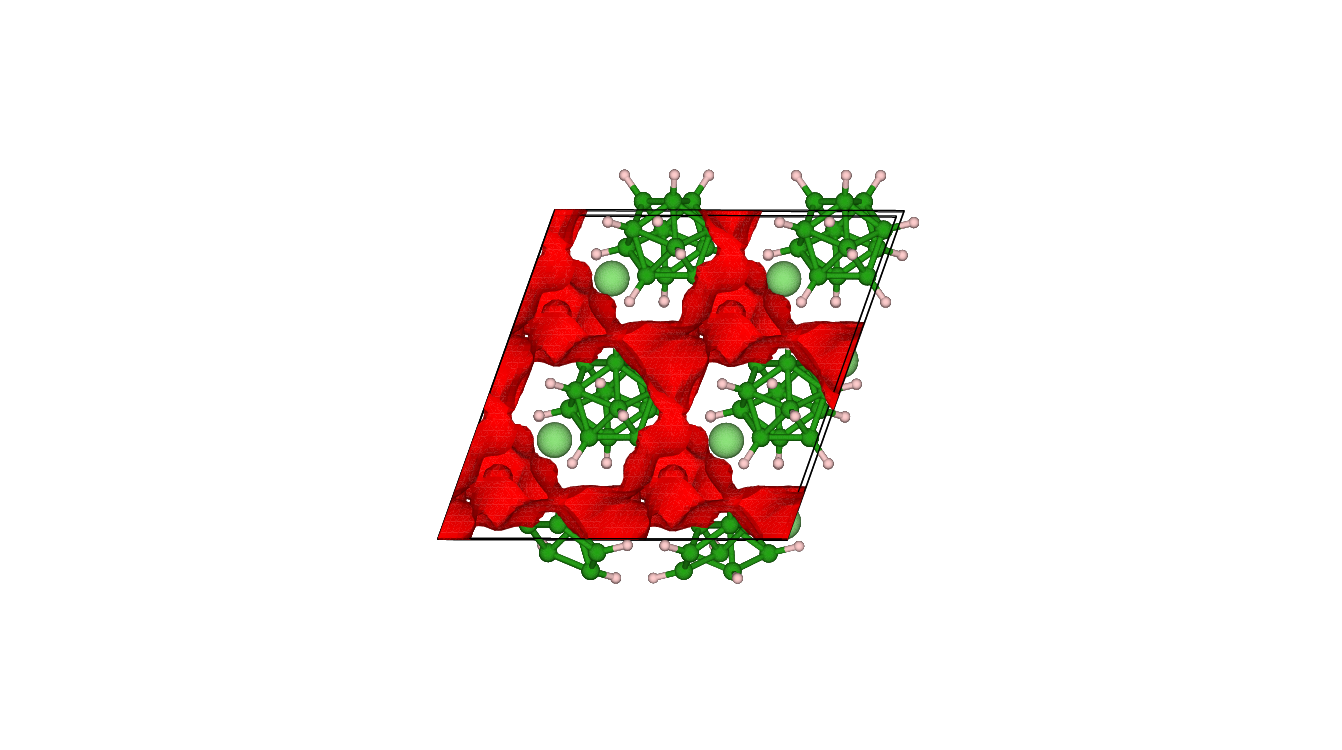}}
    }
    \hfill
    \subfloat[\label{fig:vesta:mp-2530}Li$_2$Te]{%
        \centering
        {\transpcmd\includegraphics[width=0.48\linewidth,trim={11.5cm 5.6cm 14.5cm 3.6cm},clip]{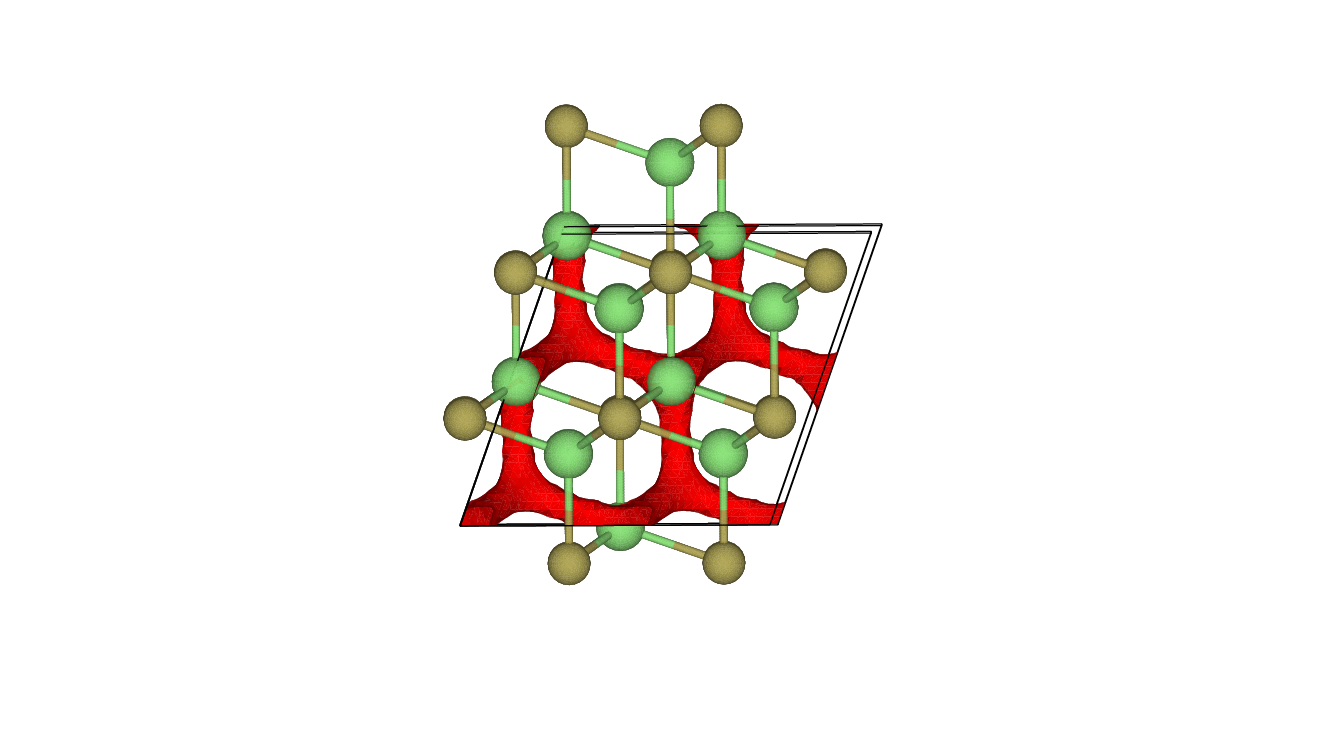}}
    }\\

    \subfloat[\label{fig:vesta:mp-1191476}Li$_2$In$_2$GeS$_6$]{%
        \centering
        {\transpcmd\includegraphics[width=0.48\linewidth,trim={12.5cm 5.6cm 13.5cm 5.6cm},clip]{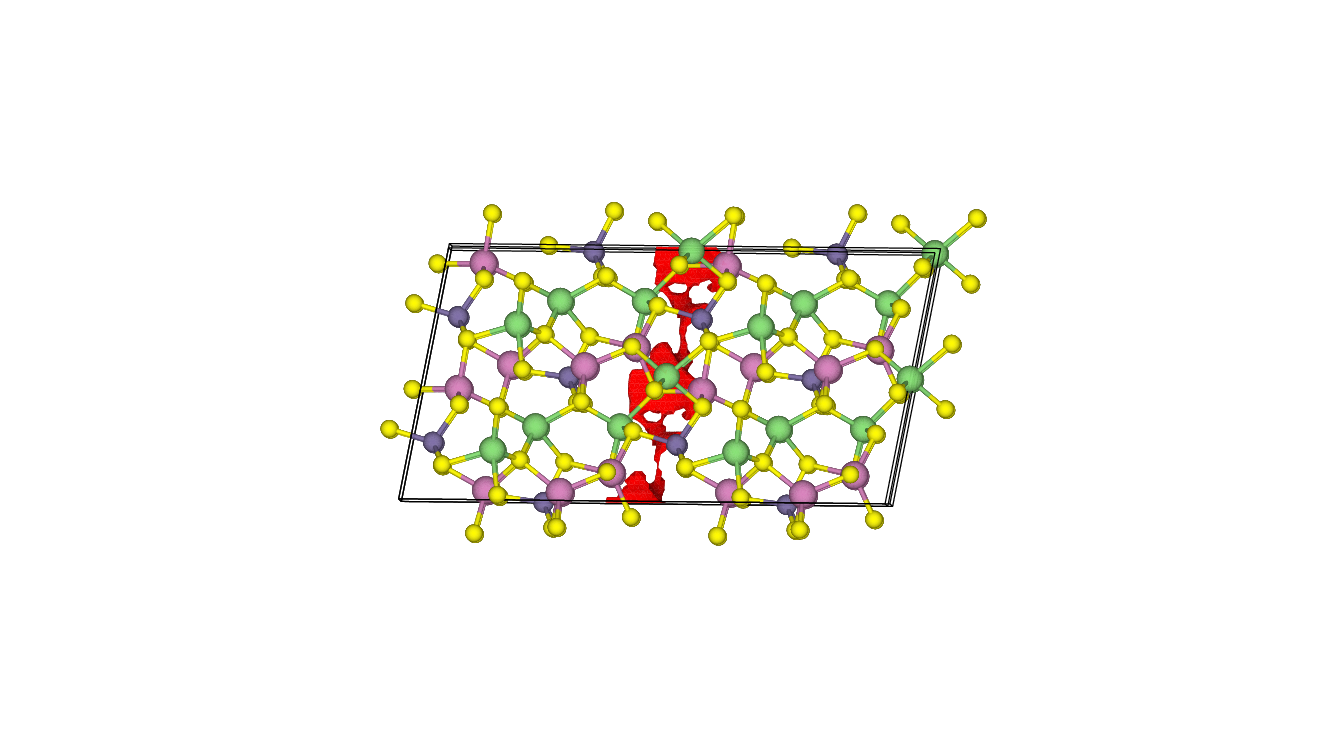}}
    }
    \hfill
    \subfloat[\label{fig:vesta:mp-768738}Li$_3$Bi(BO$_3$)$_2$]{%
        \centering
        {\transpcmd\includegraphics[width=0.48\linewidth,trim={12.5cm 5.6cm 13.5cm 5.6cm},clip]{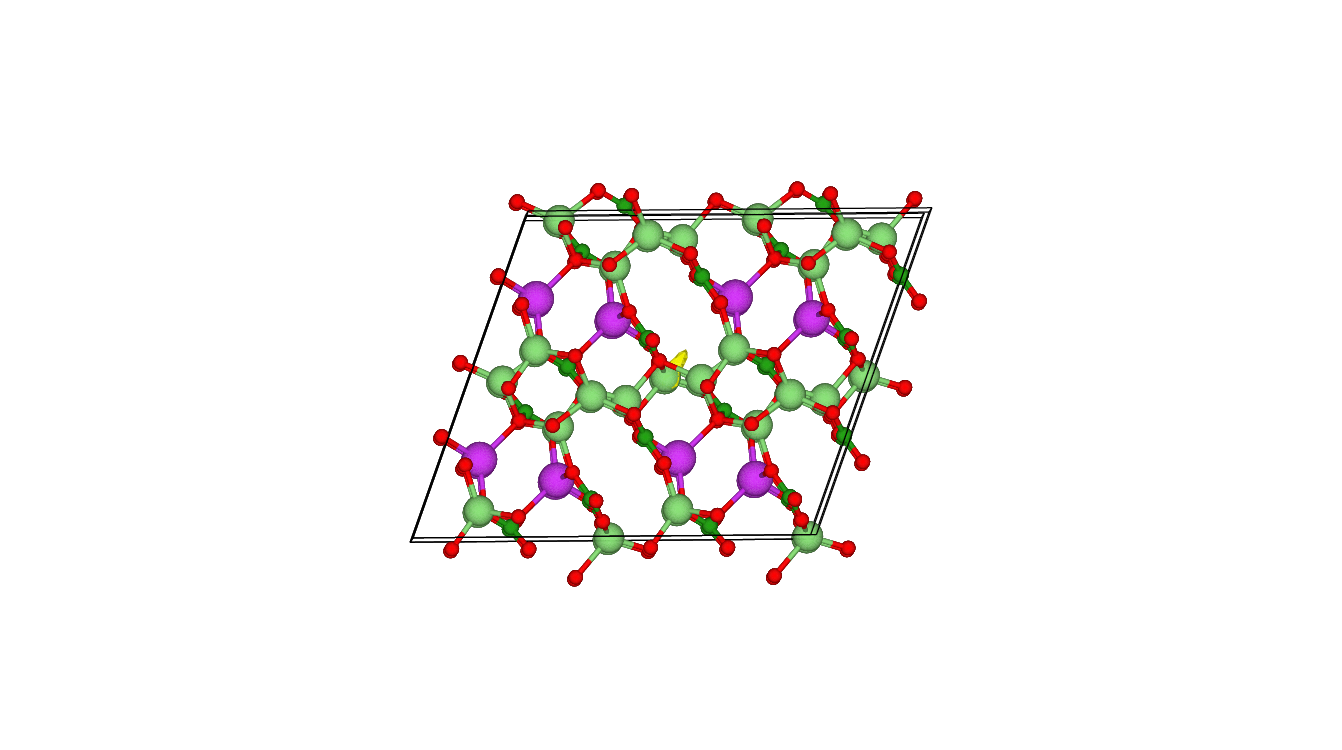}}
    }
    \caption{
        \emph{Level map isosurfaces at the threshold of 0.5\,eV.}
        The corresponding structure compositions, Materials Project identifiers and values of the combined
        $\Xi$ descriptor defined in \cref{l_met_val} (rounded to two digits) are:
        \protect\subref{fig:vesta:mp-1211296}~Li(BH)$_6$ -- \texttt{mp-1211296} -- 0.99,
        \protect\subref{fig:vesta:mp-2530}~Li$_2$Te -- \texttt{mp-2530} -- 0.97,
        \protect\subref{fig:vesta:mp-1191476}~Li$_2$In$_2$GeS$_6$ -- \texttt{mp-1191476} -- 0.35,
        and \protect\subref{fig:vesta:mp-768738}~Li$_3$Bi(BO$_3$)$_2$ -- \texttt{mp-768738} -- 0.00.
        The isosurfaces are shown in red everywhere except \protect\subref{fig:vesta:mp-768738}, where
        yellow color is used for the isosurface and red is reserved to denote the oxygen atoms.
        The structures and isosurfaces are visualized using the \textsc{VESTA} software~\cite{Momma:db5098}.
    }
    \label{fig:vesta}
\end{figure}

In order to illustrate the way our method is intended to differentiate between good and bad ionic conductors,
\cref{fig:vesta} shows the $\text{LM}(\vec{r})=0.5$\,eV isosurfaces on top of two high-FV,
one intermediate-FV and one low-FV structures. Within the frozen framework approximation, these
isosurfaces highlight the regions accessible to the central Li atom of each structure under the given
energy threshold. We expect the sizes of these regions, and hence their volumes, to correlate
with Li mobility in a given material. While the threshold of 0.5\,eV may seem extremely high, one should
anticipate that
the more accurate but expensive PES calculation with framework relaxation would have resulted in much
lower LM values.

\subsection{Validating PES descriptors}\label{l_met_val}

In order to evaluate our PES descriptors,
we make use of two publicly available datasets related to lithium mobility and conductivity.
The first dataset, \Kahle, from the study by L.\,Kahle et al.~\cite{C9EE02457C}, includes
high-throughput AIMD simulation of lithium diffusivity for 121 materials at 1000\,K,
with 25 of these materials also simulated at temperatures as low as 500\,K. The second
dataset, \Laskowski, is compiled by F.~Laskowski et al.~\cite{laskowski2023identification} in their study
focused on identifying promising ionic conductors with a semi-supervised machine learning
approach. The dataset consists of experimental conductivities from the literature,
measured at or extrapolated to room temperature. It
has 334 structures linked to the Inorganic Crystal Structure Database (ICSD)~\cite{zagorac2019recent},
75 of which can also be successfully associated with entries from the Materials Project.
To maximize the utility of the limited data, we ensure that our PES descriptors can
highlight good ionic conductors in both the simulated and experimental
datasets simultaneously. While our ultimate goal is to predict room-temperature superionic conductors,
the prevalence of high-temperature labeled structures in the \Kahle dataset is expected to improve
our method's high-temperature predictions, which is still relevant for the mentioned
ultimate goal.

The evaluation procedure is set up as follows. Firstly, since the \Kahle dataset labels the structures
with diffusion coefficients, we convert them to conductivity values via the Nernst-Einstein relation.
We also extrapolate the conductivities to room temperature for those structures in \Kahle where such
extrapolation is possible.
We then introduce two conductivity thresholds, $\sigma^{+}$ and $\sigma^{-}$, to label the structures
with the conductivity above $\sigma^{+}$ (below $\sigma^{-}$) as good (bad) conductors.
We smooth these labels with the assumption of Gaussian error distribution for the given conductivity values.
That is, we weight the samples according to the value of the integral above $\sigma^{+}$ (below $\sigma^{-}$)
for a Gaussian distribution with mean and standard deviation set to the given conductivity value and its
uncertainty.

\begin{figure}[htp]
    \centering
    {\transpcmd\includegraphics[width=\linewidth]{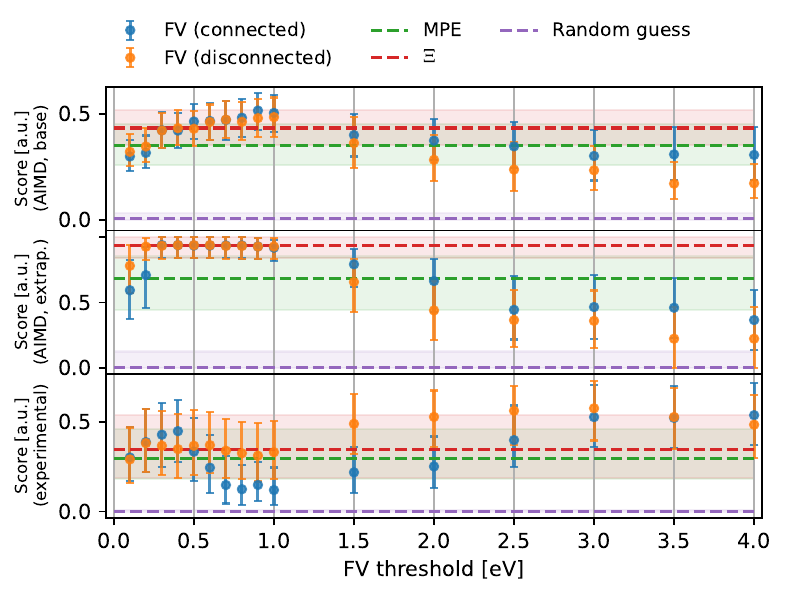}}
    \caption{\emph{Descriptor evaluation scores.}
        The top panel shows the scores obtained with $T = 1000$\,K labels from the
        \Kahle dataset. Scores obtained from that dataset with the alternative procedure,
        when positive classes are defined by the room-temperature-extrapolated conductivity
        values, are shown in the middle panel. The bottom panel contains the scores resulting
        from the experimentally measured conductivity labels reported at room temperature in
        the \Laskowski dataset. The top-performing FV descriptors, based on all three evaluation
        scores, are combined into the SSE-ranking descriptor $\Xi$ (see text) with its
        performance indicated by the dashed red line.
        For the reference, a random guess classifier performance is
        also shown (dashed purple line).
        The uncertainties are estimated with bootstrapping.
    }
    \label{fig:feature-evaluation}
\end{figure}

These labels and weights are then used to calculate ranking scores for each descriptor.
Our scoring function is a modification of the commonly used area under the ROC-curve
that only considers the lowest 10\% of the false-positive rates instead of the full integral. This ensures
that we aim towards maximal purity at the top of the list of structures
and ignore the order in the lower portion of the list
when using our descriptors for ranking potential ionic conductors.
The resulting score values for the MPE descriptor and FV descriptors at different energy thresholds are
shown in \cref{fig:feature-evaluation}.

As most of the structures from the \Kahle dataset are only simulated at $T = 1000$\,K, we
use the labels at that temperature throughout the dataset to obtain the scores shown in the top
panel of \cref{fig:feature-evaluation}, with thresholds set to
$\sigma^{+} = 0.1$\,\Scm and $\sigma^{-} = 0.01$\,\Scm. In addition, we perform a separate evaluation,
shown in the middle panel of \cref{fig:feature-evaluation},
using the room-temperature-extrapolated conductivities when determining the positive classes
with $\sigma^{+}\approx 3.16$\,\mScm (the exact relation being defined in log scale as
$\log_{10}\left(\sigma^{+} [\text{\Scm}]\right) = -2.5$). For the \Laskowski dataset, where all the
experimental conductivity
values are given at room temperature, we set the thresholds to $\sigma^{+} = 0.05$\,\mScm
and $\sigma^{-} = 0.01$\,\mScm, resulting in the scores shown in the bottom panel of
\cref{fig:feature-evaluation}.
As uncertainty values are not provided, we use an overestimate of 100\%
relative uncertainty in our weighting procedure. This, along with a lower conductivity
threshold compared to that used in the simulated data analysis, is intended to smooth out the labels
for intermediate-quality conductors.

One can see from \cref{fig:feature-evaluation} that all the proposed descriptors are capable of
predicting ionic conductors to some degree, in the sense that they all have scores higher
than a random guess. It can also be noted that FV descriptors at thresholds of $0.3-0.5$\,eV demonstrate
better or equal performance compared to MPE across all the three scores, while there is no particular
winner between the connected and disconnected versions of FV. For that reason, we pick both versions of
FV at the threshold of 0.5\,eV to be further used for ranking the structures with unknown conductivities.
We construct a combination of the
two descriptors into the SSE-ranking descriptor, denoted as $\Xi$,
by simultaneously thresholding
them with a sigmoid function:
\begin{linenomath*}\begin{equation}\label{eq:xi-definition}\begin{gathered}
    \Xi = S_1 \cdot S_2, \\
    S_1 = s_{10}\left[2.00 + \log_{10}\text{FV}^{\text{connected}}_{0.5\,\text{eV}}\right], \\
    S_2 = s_{10}\left[1.15 + \log_{10}\text{FV}^{\text{disconnected}}_{0.5\,\text{eV}}\right], \\
    s_{10}(x) \equiv\frac{1}{1+e^{-10x}}.
\end{gathered}\end{equation}\end{linenomath*}
The threshold values of $-2.00$ and $-1.15$ for the base-10 logarithms of the connected and disconnected
FV values, respectively, were picked
to simultaneously discard the low-conductivity entries in both simulated and experimental data, as
observed in \cref{fig:FV_vs_AIMD,fig:FV_vs_EXP} from Appendix~\ref{sec_fv_plots}.
The performance scores for the resulting
$\Xi$ descriptor is also shown in \cref{fig:feature-evaluation}.
Alternative ways of combining our descriptors for ranking are discussed in Appendix~\ref{appdx_alt_ranking}.

The combined descriptor is then used to rank the Materials Project structures.
We start by querying lithium containing structures with band gap of at least 0.5\,eV and energy above hull of
at most 0.05\,eV\,/\,atom, which results in 5997 potential structures. By further discarding the entries
containing transition metals we end up with 1302 structures, with 113 of them satisfying
$\Xi > 0.25$, i.e. being above the two FV thresholds simultaneously. We
then perform AIMD validation for the five of these structures with the largest $\Xi$ values.
Technical details for these simulations are given in \cref{l_met_aimd}, while the results
are presented in \cref{l_res}.

\subsection{\emph{Ab initio} molecular dynamics}\label{l_met_aimd}

Molecular dynamics simulations of selected structures with the highest $\Xi$ values
were carried out using the {\sc SIESTA} code~\cite{Soler2002}.
The forces were calculated using the generalized gradient approximation (GGA) of density functional
theory~\cite{PBE}, except for molecular solids, for which we used the LMKLL parameterization~\cite{lee2010higher}
of the van der Waals functional of Dion et al.~\cite{dion2004van}.
The core electrons are represented by pseudopotentials of the Troullier-Martins scheme~\cite{troullier1991efficient}.
The basis sets for the Kohn-Sham states are linear combinations of numerical atomic orbitals,
of the polarized double-zeta type~\cite{Sanchez-Portal1997,Sanchez-Portal2001}.

For each system, the volume relaxation was initially carried out for the primitive unit cell using
a conjugate gradient approach, and the Monkhorst and Pack Brillouin zone sampling grid~\cite{monkhorst1976special}
in the Materials Project data entry~\cite{jain2013commentary}.
A supercell capable of accommodating a sphere with a diameter of at least 6.5\,\AA\ was then constructed and the
self-consistent calculation of the charge density carried out using the $\Gamma$-point for Brillouin zone sampling.
Then, a molecular dynamics calculation for the same system was carried out at constant temperature using a Nos{\'e}
thermostat~\cite{Nose1984} and a maximum number of ten self-consistent iterations per time step.
The integration time step of 1\,fs was used. Additionally, a number of short cross-check AIMD runs with 0.25 and
0.5\,fs step sizes were carried out and showed no significant impact on the observed results.
The system is initially equilibrated in a $NVT$ ensemble for 5\,ps, after which we acquire data for the
diffusivity calculation also in the $NVT$ ensemble.

The diffusion coefficient is obtained by assuming three-dimensional Brownian motion
with a mean square displacement (MSD) given by
\begin{linenomath*}\begin{equation}
{\rm MSD} = \langle\left| \vec{r}(t)-\vec{r}(t_0)\right|^2 \rangle= 6Dt_0,
\end{equation}\end{linenomath*}
where the ensemble average is over all ions.
Since sampling trajectories for multiple starting configurations is prohibitive, we have instead
split the trajectory into multiple time windows following the
methodology used in~\cite{C9EE02457C}.
We chose the window length of 15\,ps and perform the linear fit on MSD from 1 to 15\,ps in each window.
The uncertainty for the diffusion coefficients is then estimated from the variance of these
fit results over independent fits.
A comparison of the diffusivity obtained using alternative fitting intervals can be found in
Appendix~\ref{appdx_fitting_window}.
The diffusion coefficients are converted to conductivity values using the
Nernst-Einstein relation:
\begin{linenomath*}\begin{equation}
    \sigma = \frac{nq^2}{k_{\text{B}}T}D,
\end{equation}\end{linenomath*}
where $\sigma$ is the conductivity value, $D$ is the extracted diffusion coefficient, $k_{\text{B}}$ is the
Boltzmann constant, $T$ is the absolute temperature, $n$ is the charge carrier concentration, and $q$ is
their charge.

We have also carried out tests for the finite system size effects for structure \texttt{mp-1211296}
which contains a comparatively small number of Li per supercell.
To enable access to larger supercells and extended simulation times, we employed an ML-IAP (SevenNet) to drive the
MD simulations in these tests. The results confirm that using larger supercells enhances the linearity of the MSD
dependence. Notably, the diffusion coefficients extracted from the runs with different supercell sizes
show consistent agreement within the statistical uncertainties.

\section{Results and discussion}\label{l_res}

\begin{figure}[htp]
    \centering
    {\transpcmd\includegraphics[width=\linewidth]{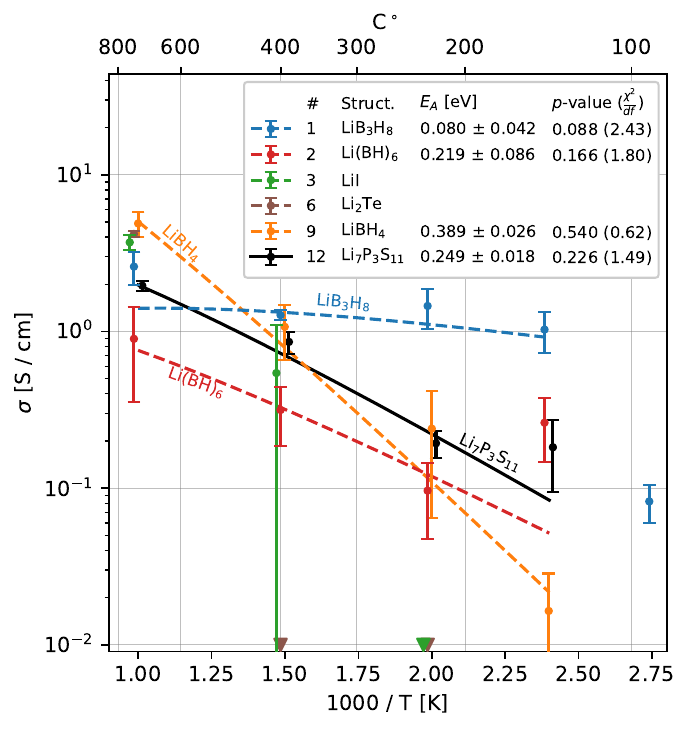}}
    \caption{\emph{AIMD conductivity values against $1 / T$ for the top-$\Xi$ structures.}
        Vertical axis is clipped at minimal conductivity value of $10^{-2}$\,\Scm with values
        below that threshold being indicated as triangles. While the target temperatures in our
        simulations
        are exactly 500, 667 and 1000\,K (with the additional simulations at 417 and 363\,K for some materials),
        small shifts are added to the horizontal axis values
        to improve marker visibility.
        The five LGPS-like structures from the top-$\Xi$ list are omitted from our AIMD
        studies and therefore not shown in this plot.
    }
    \label{fig:siestaCondVs1ovrT}
\end{figure}

We have calculated MPE and FV descriptor values, as well as the SSE-ranking descriptor $\Xi$, for
the 5997 structures from the Materials Project dataset that satisfy the minimal selection criteria, as discussed
at the end of \cref{l_met_val}. These predictions are made public and can be found at~\cite{ML4SEcode}.
We observe, that the top of the $\Xi$-ranked list is well populated by the structures from the LGPS family of
superionic conductors~\cite{C2EE23355J}, which supports the effectiveness of the proposed approach.
For the optimal use of computational resources, we therefore focus our
AIMD simulation on the other less-known structures from the top of the $\Xi$-ranked list.
Additionally, a single well-known superionic conductor,
Li$_7$P$_3$S$_{11}$~\cite{YAMANE20071163} is evaluated to validate our methodology.
The summary of our
findings is presented in \cref{tab:results}. AIMD validation results are also shown in
\cref{fig:siestaCondVs1ovrT}, where the extracted conductivity values are plotted against $1 / T$.

\begin{table*}[htp]
\newcolumntype{M}[1]{>{\centering\arraybackslash}m{#1}}
\begin{tabular}{|c|M{2.5cm}|M{2.4cm}|M{1.5cm}|M{1.25cm}M{1.25cm}M{3cm}M{2cm}M{2cm}|}
    \hline
\multirow{2}{0.4cm}{\centering \#} &
\multirow{2}{2.5cm}{\centering MP identifier~\cite{jain2013commentary}} &
\multirow{2}{2cm}{\centering Composition} &
\multirow{2}{1.75cm}{\centering $\Xi$} &
\multicolumn{5}{c|}{AIMD results \rule{0pt}{10pt}} \\
& & & & $T$ [K] & $\tau$ [ps] & $\sigma$~[S\,/\,cm] & $E_A$~[meV] & $p$-value ($\frac{\chi^2}{df}$) \\[4pt]
\hline
    1 & \texttt{mp-1211100} & LiB$_{3}$H$_{8}$ & 0.9924 & 1000 & 53.7 & $2.6 \pm 0.6$ & $80 \pm 42$ & 0.09 (2.4) \rule{0pt}{10pt}\\
     &  &  &  & 667 & 80.0 & $1.27 \pm 0.09$ &  &  \\
     &  &  &  & 500 & 104.8 & $1.5 \pm 0.4$ &  &  \\
     &  &  &  & 417 & 68.9 & $1.03 \pm 0.30$ &  &  \\
     &  &  &  & 363 & 97.4 & $0.082 \pm 0.023^*$ &  &  \\[6pt]
    2 & \texttt{mp-1211296} & Li(BH)$_{6}$ & 0.9898 & 1000 & 60.8 & $0.9 \pm 0.5$ & $219 \pm 86$ & 0.17 (1.8) \\
     &  &  &  & 667 & 81.0 & $0.32 \pm 0.13$ &  &  \\
     &  &  &  & 500 & 70.3 & $0.10 \pm 0.05$ &  &  \\
     &  &  &  & 417 & 58.4 & $0.26 \pm 0.12$ &  &  \\[6pt]
    3 & \texttt{mp-570935} & LiI & 0.9812 & 1000 & 57.2 & $3.7 \pm 0.4$ & - & - \\
     &  &  &  & 667 & 64.4 & $0.5 \pm 0.6$ &  &  \\
     &  &  &  & 500 & 66.2 & $(2.2 \pm 1.0) \times 10^{-3}$ &  &  \\[6pt]
    4 & \texttt{mp-721239} & Li$_{10}$Ge(PSe$_{6}$)$_{2}$ & 0.9746 & - & - & - & - & - \\[6pt]
    5 & \texttt{mp-721252} & Li$_{10}$Sn(PSe$_{6}$)$_{2}$ & 0.9729 & - & - & - & - & - \\[6pt]
    6 & \texttt{mp-2530} & Li$_{2}$Te & 0.9712 & 1000 & 41.9 & $4.19 \pm 0.17$ & - & - \\
     &  &  &  & 667 & 88.0 & $(1.4 \pm 0.6) \times 10^{-3}$ &  &  \\
     &  &  &  & 500 & 47.9 & $-(0.5 \pm 1.8) \times 10^{-3}$ &  &  \\[6pt]
    7 & \texttt{mp-721253} & Li$_{10}$Si(PSe$_{6}$)$_{2}$ & 0.9712 & - & - & - & - & - \\[6pt]
    8 & \texttt{mp-705516} & Li$_{10}$Sn(PSe$_{6}$)$_{2}$ & 0.9661 & - & - & - & - & - \\[6pt]
    9 & \texttt{mp-644223} & LiBH$_{4}$ & 0.9647 & 1000 & 65.6 & $4.9 \pm 0.9$ & $389 \pm 26$ & 0.54 (0.6) \\
     &  &  &  & 667 & 92.9 & $1.1 \pm 0.4$ &  &  \\
     &  &  &  & 500 & 132.7 & $0.24 \pm 0.17$ &  &  \\
     &  &  &  & 417 & 92.8 & $0.016 \pm 0.012$ &  &  \\[6pt]
    10 & \texttt{mp-696127} & Li$_{10}$Ge(PSe$_{6}$)$_{2}$ & 0.9645 & - & - & - & - & - \\[2pt]
\hline
    12 & \texttt{mp-641703} & Li$_{7}$P$_{3}$S$_{11}$ & 0.9606 & 1000 & 69.5 & $1.96 \pm 0.14$ & $249 \pm 18$ & 0.23 (1.5) \rule{0pt}{10pt}\\
     &  &  &  & 667 & 85.6 & $0.86 \pm 0.14$ &  &  \\
     &  &  &  & 500 & 119.7 & $0.193 \pm 0.039$ &  &  \\
     &  &  &  & 417 & 58.1 & $0.18 \pm 0.09$ &  &  \\[6pt]
    25 & \texttt{mp-696128} & Li$_{10}$Ge(PS$_{6}$)$_{2}$ & 0.9379 & - & - & - & - & - \\[2pt]
\hline
\end{tabular}
\caption{\emph{Top $\Xi$ predictions and AIMD results summary.}
    Ten structures with highest $\Xi$ values are shown. In addition, two known superionic conductors,
    Li$_7$P$_3$S$_{11}$~\cite{YAMANE20071163} and Li$_{10}$Ge(PS$_6$)$_2$ (LGPS)~\cite{kamaya2011lithium}, are also
    provided for reference. AIMD simulation runs were conducted with five structures from the top-ten list
    that are not obtained from LGPS by element substitution~\cite{C2EE23355J}, and also for
    one of the reference structures.
    $T$, $\tau$ and $\sigma$ represent the temperature, simulated time, and extracted conductivity value, respectively,
    for each AIMD simulation run.
    $E_A$ is the activation energy extracted by fitting the Arrhenius law for the diffusivity dependence
    on $1/T$ using weighted least squares, where applicable.
    To quantify the goodness of these fits, the $\chi^2$-statistic divided by the number of degrees of freedom
    ($\frac{\chi^2}{df}$) is reported along with the corresponding $p$-value.
    The conductivity measurement marked by the $*$ symbol was excluded from the Arrhenius law fit (see
    \cref{fig:siestaCondVs1ovrT}).
}
\label{tab:results}
\end{table*}

\subsection{AIMD validation summary}

AIMD calculations for the reference Li$_7$P$_3$S$_{11}$ structure result in the activation barrier estimate of
$E_A = 249 \pm 18$\,meV and room-temperature-extrapolated conductivity of $7.9 \pm 3.7$\,\mScm.
These are consistent with the range of values measured experimentally, with  $E_A = 295$\,meV
and $\sigma = 8.3$\,\mScm for the most pure phase sample in Ref.~\cite{busche2016situ}
and the lowest activation energy of $E_A = 124$\,meV, with
$\sigma = 3.2$\,\mScm in Ref.~\cite{mizuno2006high}.

At the top of the list, we find three hydroborate structures,
LiB$_3$H$_8$ (\texttt{mp-1211100}),  Li(BH)$_6$ (\texttt{mp-1211296}) and
LiBH$_4$ (\texttt{mp-644223}).
These belong to the family of lithium hydroborates, of which the best studied is hexagonal
LiBH$_4$~\cite{matsuo2007lithium,duchene2020status,liu2023ion}.
The latter has been found experimentally to have an ionic conductivity of the order of 1\,\mScm at 383\,K.
Cubic  Li(BH)$_6$ has been found to have low diffusivity by a combined DFT and ML-IAP study, but a closely
connected orthorhombic distortion of the structure was found to have a
conductivity of up to 0.1\,\Scm at 700\,K~\cite{maltsev2023order}.
In our AIMD calculations, all the three hydroborates show significant ionic mobility in the studied temperature range.
For LiBH$_4$, the room-temperature-extrapolated conductivity is $0.44 \pm 0.30$\,\mScm.
The other two hydroborates deviate from Arrhenius behavior at lower temperatures, preventing reliable
extrapolation of their conductivity values to room temperature.
Exclusively for LiB$_3$H$_8$, which shows very high conductivity and lowest activation energy in the temperature range
$417-1000$\,K, we perform additional AIMD simulation at $T=363$\,K yielding a conductivity value of $82\pm23$\,\mScm.
This additional measurement is not included in the Arrhenius law fit reported in \cref{tab:results} and \cref{fig:siestaCondVs1ovrT}.
Including this point yields $E_A = 216 \pm 31$\,meV with a very low $p$-value ($\frac{\chi^2}{df}$) of 0.003 (4.7).
Since these hydroborates are molecular solids, the anions have
rotational degrees of freedom, and these have been shown to
be necessary for the high cationic conductivity in Li(BH)$_6$~\cite{lu2017metal,kweon2017structural}.
Anionic motion is also apparent on the MSD of the non-Li ions (framework) in the AIMD calculations
for Li(BH)$_6$ and LiBH$_4$.

LiI (\texttt{mp-570935}) also exhibits high mobility at 1000~K,
but accompanied by observable diffusivity of the framework
ions, which may indicate that 1000\,K is close to the melting point.
At $T = 500$\,K, no diffusivity is observed.
Interestingly, in the 667\,K run, we see an abrupt change in lithium conductivity from zero during the first 35\,ps of
the simulation to over 1\,\Scm in the subsequent 30\,ps, indicating a transition to a highly diffusive state.
The value of $0.5 \pm 0.6$\,\Scm, shown in \cref{tab:results} and \cref{fig:siestaCondVs1ovrT}, is obtained by applying our MSD
analysis procedure described in \cref{l_met_aimd} to the entire run, which includes both low- and high-diffusivity segments.
Experimentally, wurtzite LiI
is metastable and has been observed~\cite{wassermann1988hexagonal}
but is very hygroscopic and hard to work with. Most ionic conductivity studies
employing LiI use cubic LiI as a minor fraction of the electrolyte mixture~\cite{poulsen1981ionic}.

Finally, Li$_2$Te (\texttt{mp-2530}) showed high diffusivity only at $T=1000$\,K, while no diffusion could be
observed at lower temperatures over the affordable simulation times.

Of the five LGPS-like structures from \cref{tab:results}, Li$_{10}$Ge(PSe$_6$)$_2$ (\texttt{mp-721239}) was
simulated in~\cite{C2EE23355J}, and the reported activation energy and conductivity at room temperature
are $190 \pm 40$\,meV and 24\,\mScm, respectively. The other structure of the same composition,
\texttt{mp-696127}, is a slightly less stable concrete realization of the same anion-substituted
disordered LGPS-like structure, and the two are expected to be mixed at room temperature.
The remaining single Li$_{10}$Si(PSe$_6$)$_2$
and the two concrete realizations of Li$_{10}$Sn(PSe$_6$)$_2$ were not
directly studied via AIMD in~\cite{C2EE23355J}, while individual substitutions of Ge $\rightarrow$
Sn/Si and S $\rightarrow$ Se were all shown to produce superionic materials.
Assuming that this also
holds for the combined substitutions, we can confirm eight of the ten top-$\Xi$ structures to be
superionic.

\subsection{ML-driven MD validation summary}

Additionally, we perform a larger-scale validation of our results using an ML-IAP as the backbone for the MD.
For this study exclusively, we employ the SevenNet model~\cite{sevennet}.
It has $\sim 4$ times more parameters and is trained on $\sim 10$ times more
data than M3GNet, which is used throughout this work, and generally outperforms it on the Matbench Discovery
benchmark. The technical details
for this SevenNet-driven validation procedure can be found in Appendix~\ref{sec_mlmd_val}.

We find that 55\% of the top-100 $\Xi$-ordered list of structures exhibit
room temperature conductivity above 0.1\,\mScm, while that proportion for an unbiased sample is only
$\sim 7$\% (2 out of 29), as predicted by SevenNet-driven MD.
Hence, we observe an approximately 8-fold increase in the number of positive examples through our selection
method. At the same time, SevenNet-driven MD predicts room temperature conductivity below
0.005\,\mScm for 39\% of the $\Xi$-selected structures.
This discrepancy in predictions between our $\Xi$-based selection
and SevenNet MD indicates that there likely are mechanisms not captured by the FV
descriptors that inhibit high ionic conductivity. It may also be attributed to the limitations of
the frozen framework approximation and potential errors in the M3GNet and SevenNet predictions.

\subsection{Discussion}

Both ML-driven MD and AIMD validation procedures confirm the significant enhancement of the amount
of room-temperature superionic materials in the sample selected using the descriptors proposed in this work.
Our method achieves even higher-purity samples if we only consider the diffusivity at 1000\,K MD simulations:
all of the AIMD-tested structures are diffusive at that temperature, and 88\% out of top-100 are diffusive
in the ML-driven MD validation. This may be due to the fact that most of the positive labels our
method was tuned on are found in the \Kahle dataset, which consists of primarily $T=1000$\,K AIMD diffusion
data. In this case, we can expect even higher performance from the proposed method once more high-quality
room-temperature conductivity data becomes available.

The highest-$\Xi$ structure we found, LiB$_3$H$_8$ (\texttt{mp-1211100}), demonstrates exceptional Li
conductivity in our AIMD studies. The SevenNet-driven MD also ranks this material with highest extrapolated
room temperature conductivity, which is an order of magnitude higher compared with the same estimate for the
reference Li$_7$P$_3$S$_{11}$ material. To the best of our knowledge, this material has not been studied
experimentally as a potential SSB electrolyte before, although the similar NaB$_3$H$_8$ structure has previously
been successfully used in a composite solid electrolyte in a sodium-metal SSB~\cite{NaB3H8battery, NaB3H8battery2}.
Ionic conductivity and relaxation times for anionic reorientation have also been studied for various phases of
the related KB$_3$H$_8$~\cite{C9DT00742C, acs.jpcc.0c10186}.

The prevalence of the hydroborate structures family at the top of the $\Xi$-ordered list, which are known
to be good ionic conductors, demonstrates the remarkable generalization capability of the proposed method.
Notably, there are no examples of positively labeled hydroborate electrolytes in either of the datasets that
we used for tuning.

We acknowledge that our descriptors cannot capture the full complexity of the dynamical effects that define
the ionic conductivity in solids~\cite{Jun2024, Fang2022, doi:10.1073/pnas.2316493121}.
However, the effectiveness observed in our screening suggests potential correlations between the complex dynamics
of ionic motion and the characteristics derived from the simplified static picture that we utilize.

Finally, we note that our proposed descriptors are also relatively fast to calculate compared to other methods
for predicting superionic materials, including ML-driven MD. Performing the
entire PES analysis on the 5997 structures from the Materials Project database took us seven days on a single
machine with 48-core Intel Xeon w7-3455 CPU and two 46-GB NVIDIA RTX 6000 Ada
Generation GPUs. For comparison, it took us approximately same time to run ML-driven MD for only 100 materials
on the same hardware.

\subsection{Outlook}

The findings presented in this study pave the way for further exploration in the field of ionic
conductors for SSBs. Firstly, while our search focused on structures within the Materials Project
database, extending this analysis to other databases, such as ICSD~\cite{zagorac2019recent}
and Alexandria~\cite{AlexandriaWebsite, AlexandriaPaper}, is straightforward and may reveal
additional promising materials.

Secondly, while the SSE-ranking
descriptor $\Xi$ was derived by manually combining the two best-performing descriptors, applying
multivariate analysis and ML techniques to the entire set of descriptors could extract even more
information from the available data and enhance prediction performance.
Moreover, we anticipate that even greater improvements will become
evident as more high-quality room-temperature conductivity data becomes available.

Another important direction is the extension of our approach to sodium conductors. We
believe that the methodology outlined in our work can be readily applied to identify promising
Na-based superionic materials. This could facilitate the advancement of sodium-ion SSBs, which
are attracting interest due to their potential for lower costs and abundant raw materials~\cite{NaASSBs}.

Finally, the pipeline we have developed is well-suited for generative modeling
applications~\cite{Al-Maeeni_2024,anonymous2024wyckoff}.
Both MPE and FV descriptors could be implemented in a differentiable manner, enabling the
calculation of gradients with respect to atomic positions and element type embeddings in the
underlying machine learning interatomic potential models. These gradients could be utilized
in a generative framework to guide the discovery of novel materials with high ionic conductivity.

\section{Conclusion}\label{l_con}

Our study presents an effective method for the rapid screening
of solid electrolyte candidates using heuristic descriptors derived from interatomic potentials.
By applying our proposed pipeline, supported by a universal machine-learned interatomic potential,
we screened 1302 compounds from the Materials Project database, identifying
several promising SSE candidates. AIMD simulations of the 10 highest-ranking
materials confirm that all are conductive at high temperatures, with 8 out of 10 being
superionic at room temperature. Notably, our approach highlights the solid electrolyte
LiB$_3$H$_8$ (\texttt{mp-1211100}), with an impressive AIMD ionic conductivity estimate of
$82 \pm 23$\,\mScm at $T=363$\,K. To our knowledge, this material has not been
studied experimentally in the context of SSBs.

\section{Data Availability Statement}
The code implementing the proposed technique, as well as the predictions calculated for lithium-containing
structures from the Materials Project, are available in~\cite{ML4SEcode}.

\section{Acknowledgements}
This research project is supported by the Ministry of Education,
Singapore, under its Research Centre of Excellence award to the
Institute for Functional Intelligent Materials, National University of Singapore
(I-FIM, project No. EDUNC-33-18-279-V12).
This research/project is supported by the National Research Foundation, Singapore under its AI
Singapore Programme (AISG Award No: AISG3-RP-2022-028).
This work used computational resources of the Singapore National Supercomputing Centre (NSCC) of
Singapore. For the exploratory phase, this work used computational resources of the Constructor
Research Platform provided by Constructor Technologies.

\bibliography{refs}

\clearpage
\begin{appendices}
    \section{Validation with ML-driven MD}\label{sec_mlmd_val}

\begin{figure}[htp]
        \centering
        {\transpcmd\includegraphics[width=\linewidth]{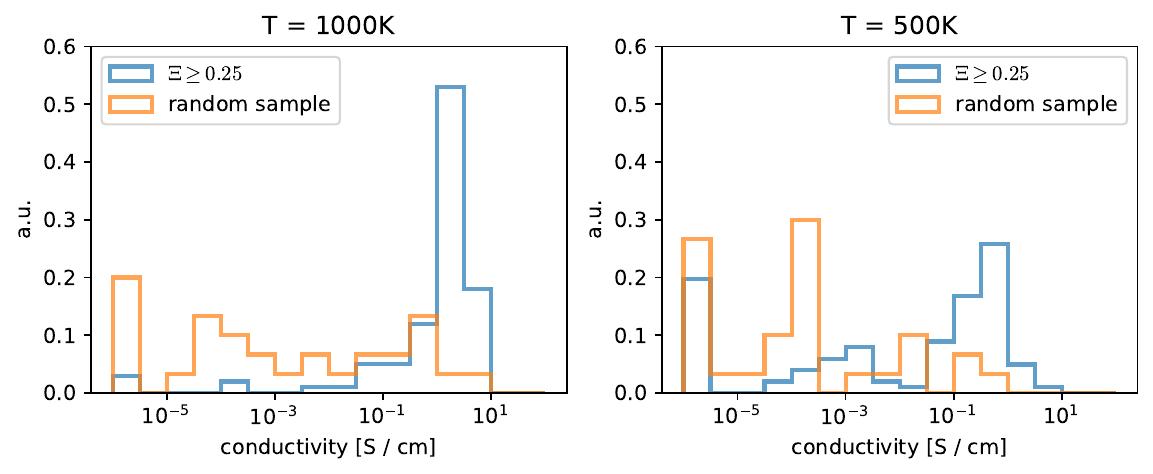}} \\
        {\transpcmd\includegraphics[width=0.7\linewidth]{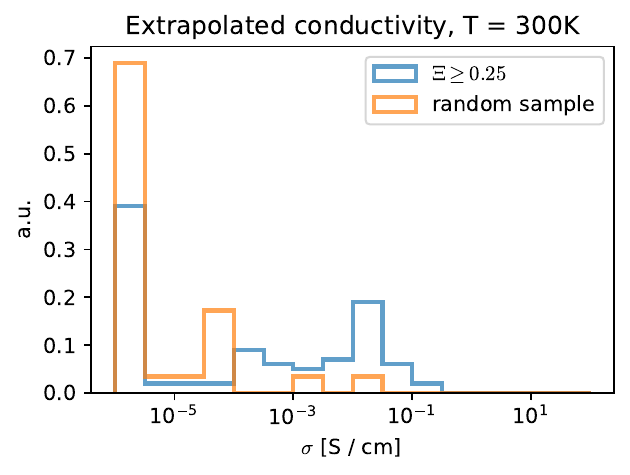}}
        \caption{\emph{Distributions of conductivities estimated with MD driven by SevenNet.}
            A random sample of size 30 is compared with 100 structures from the top of the
            $\Xi$-ordered list. In each panel, both histograms are normalized to have the same
            total weight. The top left (right) panel shows the distributions of values
            extracted from the $T = 1000$\,K ($T = 500$\,K) runs. The bottom panel shows
            the distributions obtained by extrapolating to the room temperature.
        }
        \label{fig:sevennet}
    \end{figure}

\begin{figure}[htp]
        \centering
        {\transpcmd\includegraphics[width=\linewidth]{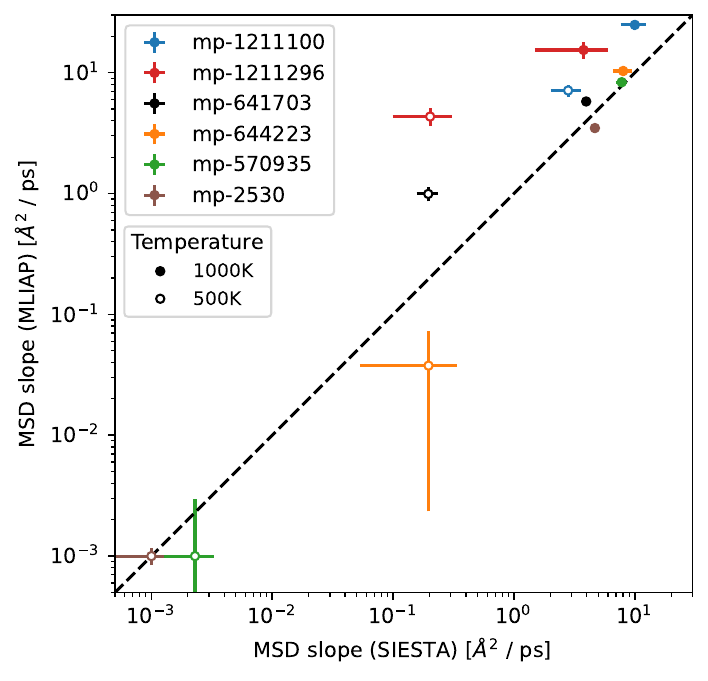}}
        \caption{\emph{MSD slopes comparison between AIMD and SevenNet-driven MD.}
            All values are clipped from below at $10^{-3}$\,$\AA^2$\,/\,ps.
            The dashed black line denotes the set of points with $x = y$.
        }
        \label{fig:sevennet_vs_aimd}
    \end{figure}

    We perform an additional validation procedure for our predictions using MD simulations driven
    by an ML-IAP, which allows to significantly scale up the number of validated structures, as
    compared to the AIMD studies. While all our PES descriptors are calculated with
    the M3GNet model~\cite{chen2022universal}, we choose the newer SevenNet model~\cite{sevennet},
    which outperforms M3GNet on the Matbench Discovery benchmark~\cite{riebesell2023matbench}.
    These MD simulations are carried out on 30 random structures satisfying the minimal selection
    described in \cref{l_met_val}, as well as on the top 100 structures ordered by the $\Xi$ values.
    Each structure is simulated at two temperature points of 1000\,K and 500\,K, with step size
    of 1\,fs and total simulation time of 100\,ps for each run. Then, the same procedure as
    for the AIMD studies is used to extract the conductivity values. \Cref{fig:sevennet}
    shows the distributions of the obtained conductivity values. It can be clearly seen from
    these plots that $\Xi$-based selection significantly enhances the sample with structures
    with high conductivity at both simulated temperature points and at room temperature, the latter
    being assessed by extrapolation. In order to demonstrate the level of agreement between AIMD and SevenNet,
    we plot the two sets of the extracted MSD slope values against each other in
    \cref{fig:sevennet_vs_aimd}. Finally, we present the structures that SevenNet confirmed to be superionic,
    excluding those previously reported in~\cite{sendek2018machine,laskowski2023identification,muy2019high},
    in \cref{tab:predictions}.

    \begin{table}[htp]
    \begin{tabular}{lllll}
        \multicolumn{1}{l}{MP identifier\hspace{0.1cm}~} &
        \multicolumn{1}{l}{Composition\hspace{0.2cm}~} &
        \multicolumn{2}{l}{System (space group)\hspace{0.1cm}~} &
        \multicolumn{1}{c}{$\Xi$} \\[2pt]
        \hline
        \texttt{mp-1211100} & LiB$_3$H$_8$                   &  Orthorhombic &  (63)   & 0.992 \rule{0pt}{10pt}\\
        \texttt{mp-1211296} & Li(BH)$_{6}$                   &         Cubic &  (202)  & 0.99  \\
        \texttt{mp-985583}  & Li$_3$PS$_4$                   &  Orthorhombic &  (62)   & 0.958 \\
        \texttt{mp-1001069} & Li$_{48}$P$_{16}$S$_{61}$      &    Monoclinic &  (6)    & 0.956 \\
        \texttt{mp-1097034} & Li$_{20}$Si$_3$P$_3$S$_{23}$Cl &    Monoclinic &  (6)    & 0.951 \\
        \texttt{mp-1040451} & Li$_{20}$Si$_3$P$_3$S$_{23}$Cl &    Monoclinic &  (6)    & 0.949 \\
        \texttt{mp-1185263} & LiPaO$_3$                      &         Cubic &  (221)  & 0.942 \\
        \texttt{mp-1097036} & Li$_3$PS$_4$                   &  Orthorhombic &  (62)   & 0.931 \\
        \texttt{mp-1222398} & LiGa(GeSe$_3$)$_2$             &    Monoclinic &  (9)    & 0.896 \\
        \texttt{mp-755463}  & Li$_3$SbS$_3$                  &      Trigonal &  (148)  & 0.887 \\
        \texttt{mp-753720}  & Li$_3$BiS$_3$                  &      Trigonal &  (148)  & 0.858 \\
        \texttt{mp-1222392} & LiUI$_6$                       &    Monoclinic &  (5)    & 0.82  \\
        \texttt{mp-34038}   & Li$_6$NCl$_3$                  &    Monoclinic &  (8)    & 0.814 \\
        \texttt{mp-680395}  & Li$_3$As$_7$                   &  Orthorhombic &  (61)   & 0.783 \\
        \texttt{mp-775806}  & Li$_3$SbS$_3$                  &    Monoclinic &  (14)   & 0.763 \\
        \texttt{mp-28336}   & Li$_3$P$_7$                    &  Orthorhombic &  (19)   & 0.721 \\
        \texttt{mp-1222582} & Li$_4$GeS$_4$                  &  Orthorhombic &  (33)   & 0.717 \\
        \texttt{mp-753429}  & Li$_4$Bi$_2$S$_7$              &    Monoclinic &  (14)   & 0.716 \\
        \texttt{mp-985582}  & Li$_6$PS$_5$I                  &         Cubic &  (216)  & 0.703 \\
        \texttt{mp-1177520} & Li$_3$SbS$_3$                  &      Trigonal &  (160)  & 0.658 \\
        \texttt{mp-1211362} & Li(BH)$_6$                     &    Monoclinic &  (14)   & 0.647 \\
        \texttt{mp-1211446} & Li$_7$PSe$_6$                  &  Orthorhombic &  (33)   & 0.606 \\
        \texttt{mp-1222482} & Li$_6$AsS$_5$I                 &     Triclinic &  (1)    & 0.539 \\
        \texttt{mp-1211324} & Li$_7$PS$_6$                   &  Orthorhombic &  (33)   & 0.538 \\
        \texttt{mp-1195718} & Li$_4$SnS$_4$                  &  Orthorhombic &  (62)   & 0.525 \\
        \texttt{mp-950995}  & Li$_6$PS$_5$I                  &    Monoclinic &  (9)    & 0.521 \\
        \texttt{mp-1211176} & Li$_6$AsS$_5$I                 &    Monoclinic &  (9)    & 0.483 \\
    \end{tabular}
    \caption{\emph{Top-$\Xi$ candidates confirmed with SevenNet MD that were not predicted
        in~\cite{sendek2018machine,laskowski2023identification,muy2019high}.}
        Reported are the 27 structures (25 after excluding the two containing rare and radioactive elements)
        from the top-100 highest $\Xi$ list that have extrapolated
        room temperature conductivity above 1\,\mScm, as predicted with SevenNet MD.
        Structures already reported in~\cite{sendek2018machine,laskowski2023identification,muy2019high}
        were excluded from this list.
    }
    \label{tab:predictions}
    \end{table}

    \section{Comparison of FV descriptors with AIMD and experimental labels}\label{sec_fv_plots}

\begin{figure*}[htp]
        \centering
        \subfloat{%
            {\transpcmd\includegraphics[width=0.44\linewidth]{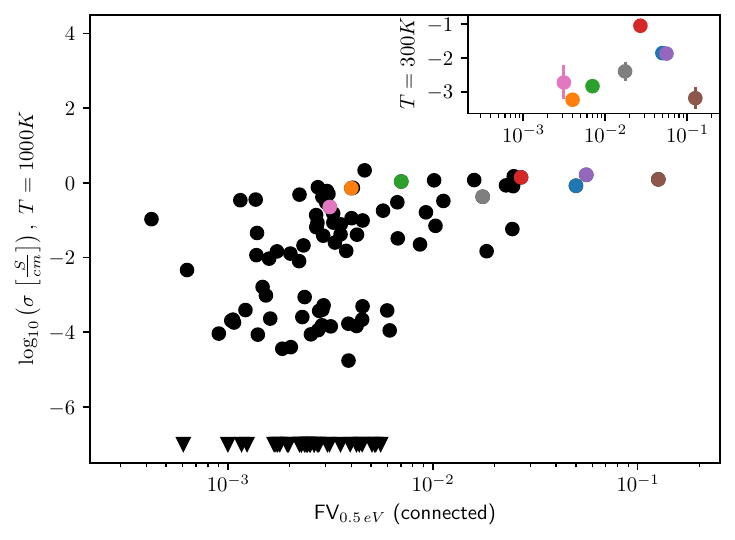}}
        }
        \subfloat{%
            {\transpcmd\includegraphics[width=0.44\linewidth]{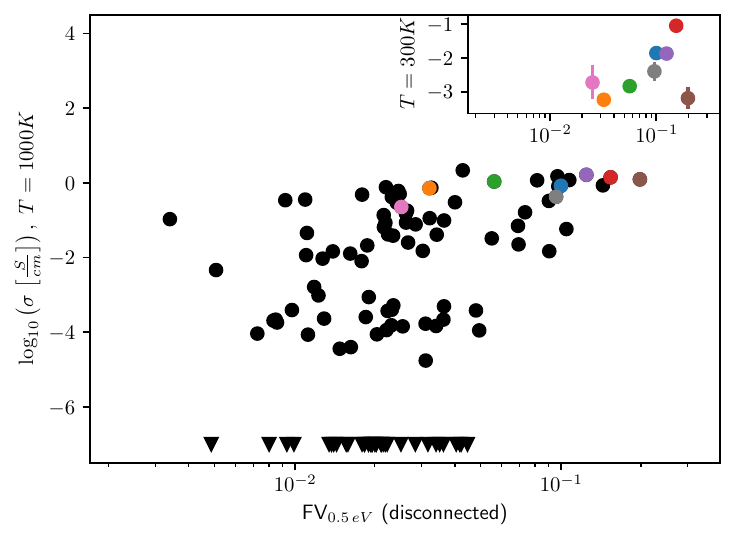}}
        }
        \caption{\emph{Distributions of conductivity labels against FV descriptors for the simulated dataset.}
            Shown are such distributions for the connected (left) and disconnected (right) versions of the
            FV descriptors at the threshold of 0.5\,eV. The conductivity labels are taken from the 1000\,K
            temperature point in the main plots, while the inset plots show such distributions for the
            room-temperature-extrapolated conductivity values. As the extrapolated temperature entries are
            a small subset of the whole dataset, they use per-structure colors in order to relate same
            structures between the two temperature points. Since the method for extracting the diffusivity
            constants used in~\cite{C9EE02457C}
            allows for small negative values, such values are clipped
            at $\sigma = 10^{-7}$\,\Scm to be kept in the log-scale and shown as triangles in the plots.
        }
        \label{fig:FV_vs_AIMD}
    \end{figure*}

\begin{figure*}[htp]
        \centering
        \subfloat{%
            {\transpcmd\includegraphics[width=0.4\linewidth]{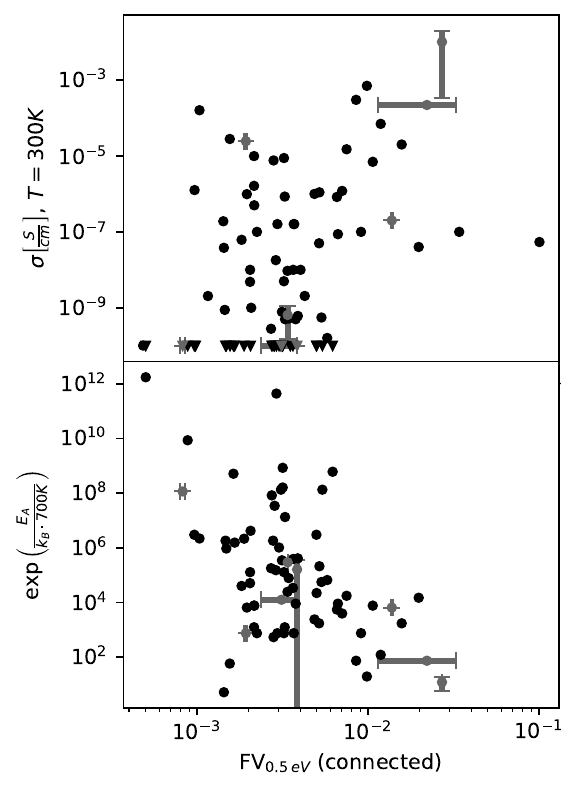}}
        }
        \subfloat{%
            {\transpcmd\includegraphics[width=0.4\linewidth]{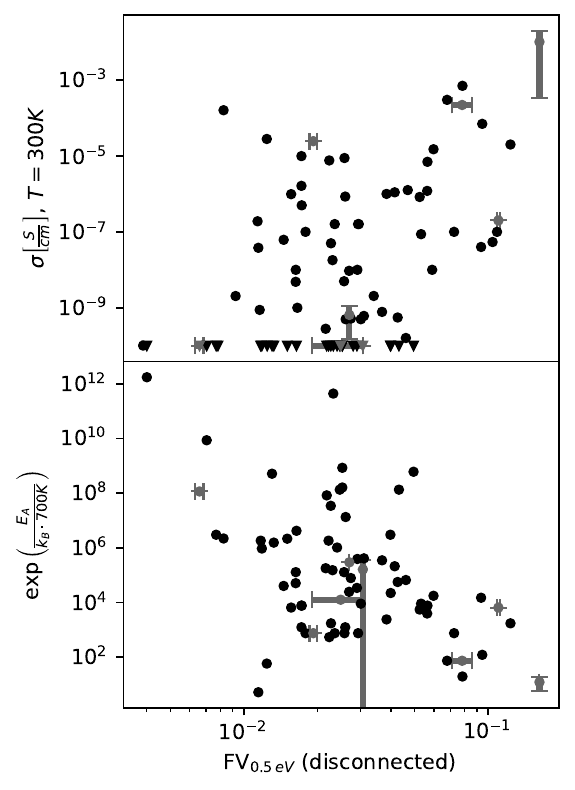}}
        }
        \caption{\emph{Distributions of conductivity and activation energy labels against FV descriptors
            for the experimental dataset.} Shown are such distributions for the connected (left) and
            disconnected (right) versions of the FV descriptors at the threshold of 0.5\,eV. The top
            panels show the distributions for the conductivities at room temperature. Extremely low
            conductivity values are clipped at $\sigma = 10^{-10}$\,\Scm and are shown as triangles
            in the plot. The bottom panels demonstrate the activation energies provided in the
            \Laskowski dataset converted to a multiplicative factor relating diffusion constants
            between $T = 1000$\,K and $T = 300$\,K via the Arrhenius law.
            Horizontal error bars are provided for the structures taken from ICSD, where
            partial occupancy sites are present and a number of concrete structures are sampled.
            Vertical error bars are shown for the cases where multiple measurements are provided for a
            single structure.
        }
        \label{fig:FV_vs_EXP}
    \end{figure*}

    \Cref{fig:FV_vs_AIMD,fig:FV_vs_EXP}
    demonstrate scatter plots with relationships
    between the labels from the \Kahle and \Laskowski datasets, respectively, and the connected and
    disconnected versions of the FV descriptor at the threshold of 0.5\,eV.


    \section{Alternative ranking models}\label{appdx_alt_ranking}
    The ranking descriptor $\Xi$ from \cref{eq:xi-definition} was originally constructed through a visual
    examination of \cref{fig:feature-evaluation,fig:FV_vs_AIMD,fig:FV_vs_EXP}.
    In this section, we describe our work to develop an improved ranking model, using a multivariate analysis
    of all the introduced descriptors (MPE and FV), referred to collectively as the \emph{PES descriptors}.
    Additionally, we incorporate the descriptors from the \texttt{matminer} library~\cite{WARD201860},
    following~\cite{laskowski2023identification}, to evaluate the relative performance of the two descriptor
    sets, as well as their combined utility.

    \begin{figure}[htp]
        \centering
        \vspace{10pt}
        {\transpcmd\includegraphics[width=\linewidth]{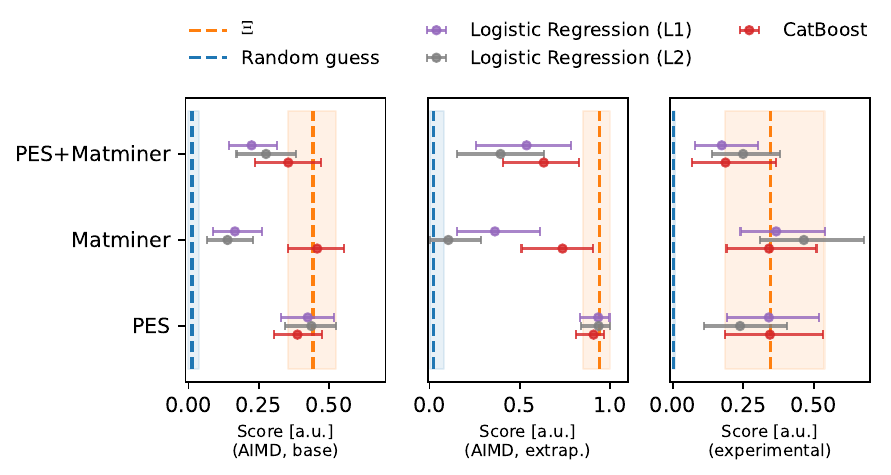}}
        \caption{\emph{Evaluation scores for models trained on the \Kahle dataset using
            various sets of descriptors.}
            The scoring procedure is the same as for the results presented in \cref{fig:feature-evaluation}.
        }
        \label{fig:matminercomp}
    \end{figure}

    To make this comparison, we train classification models on the \Kahle dataset with $T=1000$\,K labels, as
    defined in \cref{l_met_val}, and evaluate them using the procedure outlined in that section.
    The Leave-One-Out cross-validation technique was employed for model training, where each individual sample is
    sequentially left out for testing while the remaining samples are used for training.
    The following three multivariate analysis models were trained: two versions of logistic regression, with L1
    and L2 regularization terms, respectively, as implemented in~\cite{scikit-learn}, as well as the
    CatBoost~\cite{dorogush2018catboost} implementation of gradient boosting on decision trees.
    Each model was separately trained on three feature representations: based on PES descriptors alone, based
    on \texttt{matminer} descriptors alone, as well as on the combination of both.
    Resulting scores are shown in \cref{fig:matminercomp}.

    From \cref{fig:matminercomp}, it appears that none of the models outperform the $\Xi$ descriptor.
    Models trained exclusively on the \texttt{matminer} descriptors show reasonable scores on the
    \Laskowski experimental dataset, but they sacrifice predictive power on the AIMD data.

    \section{Alternative MSD fitting window parameters}\label{appdx_fitting_window}
\begin{figure*}[htp]
        \centering
        {\transpcmd\includegraphics[width=0.7\linewidth]{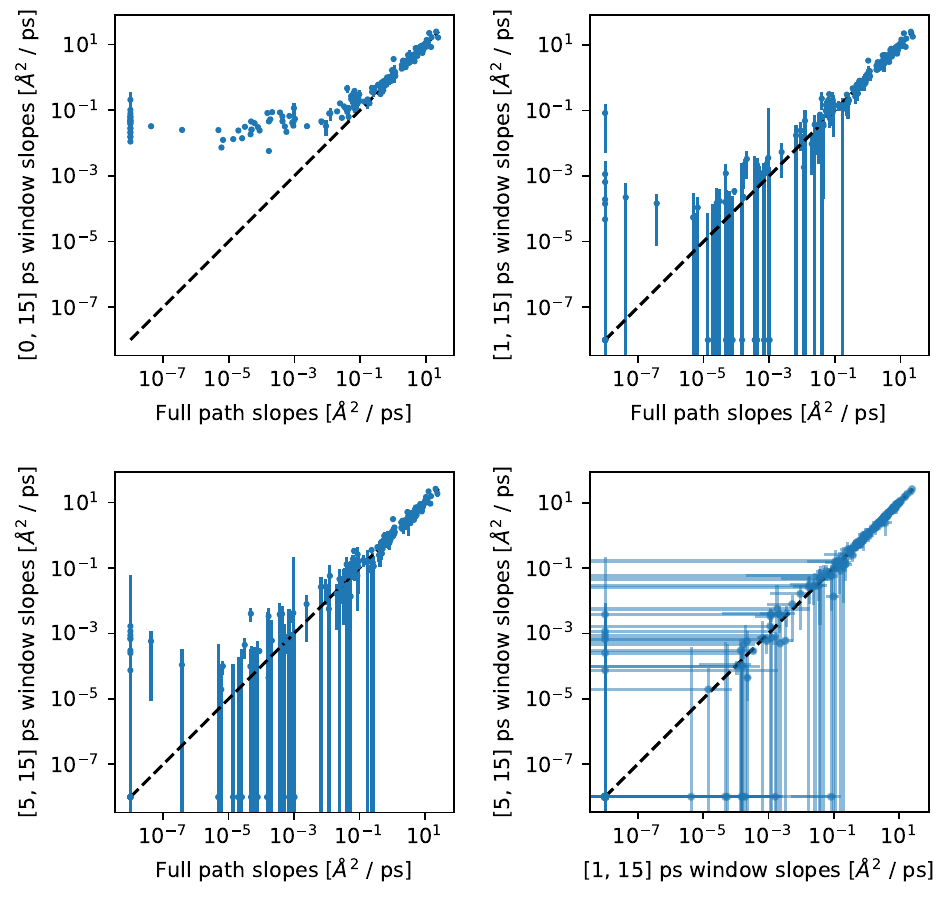}}
        \caption{\emph{MSD slope values extracted with different fitting window parameters.}
            The results of single fits for the entire simulated trajectory (horizontal axes) are compared with
            various windowed fit configurations (vertical axes) in top left, top right and bottom left panels,
            while bottom right panel compares two windowed fit configurations against each other.
            All windowed fits are performed for 15\,ps windows.
            The starting point of the windowed fits is 0\,ps in the top left panel, 1\,ps in the top right panel and
            5\,ps in the bottom left panel. Bottom right panel compares 1\,ps and 5\,ps fit starting points.
            The values are extracted from all the SevenNet MD and AIMD runs performed.
            All values are clipped from below at $10^{-8}$\,$\AA^2$\,/\,ps.
            The dashed black line denotes the set of points with $x = y$.
        }
        \label{fig:fittingwin}
    \end{figure*}

    Along with the MSD fitting procedure described in the main text, we consider three alternative options. The first
    option is to fit the entire MSD trajectory, which doesn't allow to properly estimate the uncertainty of the result
    but is most sensitive to slowest diffusion processes. The two remaining options are alternative windowed
    configurations with fitted MSD interval starting from 0\,ps and 5\,ps, respectively, as opposed to 1\,ps
    chosen in the main procedure. The comparison of the results obtained with the main and the three alternative
    procedures is shown in \cref{fig:fittingwin}. We can clearly see that all four approaches agree in
    high diffusion regime. In the slow diffusion region, fitting each window from 0\,ps overestimates the slopes of
    MSD due to periodic atomic movement and therefore is inapplicable. Windowed fits with later starting point of 5\,ps
    produce results that are consistent with the 1\,ps starting point fits.

\end{appendices}

\ifincludeinternal
    \newpage
    \include{internal}
\fi

\end{document}

%% file: common_commands.tex
\usepackage{xspace}
\newcommand{\Kahle}{{\textsc{Kahle2020}}\xspace}
\newcommand{\Laskowski}{{\textsc{Laskowski2023}}\xspace}
\newcommand{\Scm}{{S\,/\,cm}\xspace}
\newcommand{\mScm}{{mS\,/\,cm}\xspace}